\documentclass[12pt,eqsecnum]{article}
\usepackage[dvips]{graphicx}
\usepackage{amssymb}
\usepackage{amsmath}
\usepackage{epstopdf}
\usepackage{color}
\usepackage{soul}
\usepackage{ulem}
\makeatletter

\@addtoreset{equation}{section}
\makeatletter

\newcommand{\D}{{\rm d}}
\newcommand{\dalm}{\kern1pt\vbox{\hrule height 0.9pt\hbox{\vrule width
0.9pt\hskip 2.5pt\vbox{\vskip 5.5pt}\hskip 3pt\vrule width 0.3pt}\hrule height
0.3pt}\kern1pt}

\begin{document}

\begin{titlepage}
\vfill
\begin{flushright}
\today
\end{flushright}

\vfill
\begin{center}
\baselineskip=16pt
{\Large\bf 
All static and electrically charged solutions with Einstein base manifold in the arbitrary-dimensional Einstein-Maxwell system with a massless scalar field\\
}
\vskip 0.5cm
{\large {\sl }}
\vskip 10.mm
{\bf  Hideki Maeda${}^{a}$ and Cristi{\'a}n Mart\'{\i}nez$^{b}$} \\

\vskip 1cm
{
   	${}^a$ Department of Electronics and Information Engineering, Hokkai-Gakuen University, Sapporo 062-8605, Japan.\\
    ${}^b$ Centro de Estudios Cient\'{\i}ficos (CECs), Av. Arturo Prat 514, Valdivia, Chile. \\
	\texttt{h-maeda@hgu.jp, martinez@cecs.cl}

     }
\vspace{6pt}
\end{center}
\vskip 0.2in
\par
\begin{center}
{\bf Abstract}
\end{center}
\begin{quote}
We present a simple and complete classification of static solutions in the Einstein-Maxwell system with a massless scalar field in arbitrary $n(\ge 3)$ dimensions.
We consider spacetimes which correspond to a warped product $M^2 \times K^{n-2}$, where $K^{n-2}$ is a $(n-2)$-dimensional Einstein space. The scalar field is assumed to depend only on the radial coordinate and the electromagnetic field is purely electric. Suitable \textit{Ans{\"a}tze} enable us to integrate the field equations in a general form and express the solutions  in terms of elementary functions. The classification with a non-constant real scalar field consists of nine solutions for $n\ge 4$ and three solutions for $n=3$. 
A complete geometric analysis of the solutions is presented and the global mass and electric charge are determined for  asymptotically flat configurations. 
There are two remarkable features for the solutions with $n\ge 4$: (i) Unlike the case with a vanishing electromagnetic field or constant scalar field, asymptotically flat solution is not unique, and (ii) The solutions  can asymptotically approach the Bertotti-Robinson spacetime depending on the integrations constants.
In accordance with the no-hair theorem, none of the solutions are endowed of a Killing horizon. 
\vfill
\vskip 2.mm
\end{quote}
\end{titlepage}



\tableofcontents

\newpage

\section{Introduction}

The role of exact solutions in physics is to understand the properties of physical phenomena in a variety of situations.
Even a particular solution may play a large role for our understanding of nature.
In gravitational physics, good examples of such exact solutions are Schwarzschild and Kerr solutions.
By the black hole uniqueness theorem, asymptotically flat static black holes in vacuum are represented by Schwarzschild solution \cite{Israel1967} and this result has been extended to the stationary case in which Kerr solution represents the unique black hole \cite{Carter1971}.
This implies that, in the asymptotically flat and stationary spacetime in vacuum, all the properties of a black hole are encoded in the Kerr black hole and it lead us to the discovery of the black-hole mechanics~\cite{BHmechanics} and then its celebrated thermodynamics description \cite{BHthermodynamics}.

However, this strong uniqueness result does not mean that the final state of gravitational collapse is always a Kerr black hole because there might be other solutions which do not represent a black hole but a star or naked singularity.
In this context, the cosmic censorship hypothesis has been proposed, which asserts that the final state in physically reasonable and generic situations cannot be a naked singularity~\cite{penrose1969,penrose1979}.
Although the generic proof of this hypothesis is far from complete, it has been studied in the systems with symmetry which makes the problem tractable.
Among others, in the spherically symmetric spacetime, the cosmic censorship hypothesis was shown to be false, where a naked singularity is formed in the dynamical region with matter \cite{christodoulou1984}.
On the other hand, it was shown to be true in the case with a massless scalar field \cite{ns-scalar}.

These results suggest that the final static configuration after the collapse may be not a Schwarzschild black hole but a naked singularity.
Then, a complete classification of all the spherically symmetric and static solutions must be quite helpful for future investigations to find the candidates of the final configuration.
Actually, it was shown~\cite{JNWhigher} that the general spherically symmetric and static solution for a massless scalar field is the so-called Janis-Newman-Winicour (JNW) solution\footnote{Actually, this solution was first obtained by Fisher~\cite{Fisher1948} and rediscovered many times by different authors~\cite{jnw1968,bl1957,Buchdahl1959,wyman1981}.} \cite{jnw1968} which contains one additional parameter to the mass parameter in the Schwarzschild solution.

The previous analysis can be naturally extended in the presence of the Maxwell field in addition.
The unique static black hole in this system is the Reissner-Nordstr\"om black hole \cite{Israel1968} and the exterior of this charged black hole is stable against linear perturbations \cite{RNBHstability} similar to the Schwarzschild black hole \cite{SchBHstability}.
However, it suffers from different kinds of instability.
One is the mass-inflation instability of the inner horizon \cite{PissonIsrael1990} which transform a part of the inner horizon into a curvature singularity in gravitational collapse of a massless scalar field with the Maxwell field \cite{Dafermos2003}.
It was also found that extreme Reissner-Nordstr\"om black hole suffers from a different type of instability at the extremal horizon where the second derivative of a massless scalar field generically grows with time \cite{aretakis2011,mrt2013}. 

This class of problems motivate us to perform a complete classification of all the spherically symmetric and static solutions with a massless scalar field and purely electric Maxwell field.
After the earlier works in~\cite{Penney1969,Bronnikov1973}, all the asymptotically flat and static solutions have been certainly classified in four dimensions in~\cite{vdb1983}, however, the metric functions and the scalar field are given in terms of the electric potential.
Results in arbitrary dimensions were presented in \cite{pk1996}, where a spherically symmetric reduction to a two-dimensional dilatonic-gravity theory coupled with a $U(1)$ gauge field was considered. However, in the latter work it seems difficult to see explicit forms of all the solutions, which is a requirement for describing their physical properties.

Actually in the neutral case, the JNW solution, namely the general static and spherically symmetric solution in four dimensions, can be obtained by certain solution-generating methods from the Schwarzschild solution~\cite{Buchdahl1959,AbdolrahimiShoom2010}.
Similarly, an electrically charged solution found by Penney~\cite{Penney1969} can be obtained by a different solution-generating method~\cite{jrw1969} from the JNW solution~\cite{mm-preparation}.
However, this does not mean that the Penney solution is the general solution.
In fact, the general static and spherically symmetric solution is not unique in the charged case.

 In the present paper, we present {\it all} static solutions in a closed form in a more general setup, which includes $(n-2)$-dimensional Einstein spaces as base manifolds in $n(\ge 3)$ arbitrary spacetime dimensions. The key point for obtaining this complete classification is the choice of an adequate coordinate system which allows us to integrate the field equations in a direct and transparent way, leading to simple expressions for the solutions.

In the next section we introduce the action and the corresponding field equations. From the general expressions for a static metric and  for a  radial scalar and electric field, it is shown the absence of a Killing horizon in presence a non-constant scalar field, in agreement with the no-hair theorem. In Sec. 3, the field equations are solved for four and higher dimensions and the complete set of the solutions is expressed in a very simple closed form. The three-dimensional case is also solved in Sec. 4.  In both sections a detailed analysis of the geometrical properties of the solutions, including singularities and limiting cases, is exhibited.  In Sec. 5, we identify solutions which represent asymptotically flat spacetimes and the mass and electric charge of those configurations are determined. Here we show two important properties of the solutions  in  $n\ge 4$ dimensions: (i) As opposite  the case with a vanishing electromagnetic field or constant scalar field, the asymptotically flat solution is not unique, and (ii) The solutions  can asymptotically approach the Bertotti-Robinson spacetime depending on the integrations constants.
After some concluding remarks, an appendix is included. 
Appendix A contains a simple derivation of the full set of solutions, which is obtained by using an alternative radial coordinate.

Our basic notations follow~\cite{wald}.
The conventions of curvature tensors are 
$[\nabla _\rho ,\nabla_\sigma]V^\mu ={\cal R^\mu }_{\nu\rho\sigma}V^\nu$ 
and ${\cal R}_{\mu \nu }={\cal R^\rho }_{\mu \rho \nu }$.
The Minkowski metric is taken to be the mostly plus sign, and 
Roman indices run over all spacetime indices. 
The $n$-dimensional gravitational
constant is denoted by  $\kappa_{n}$, and the electromagnetic field strength is given by $F_{\mu\nu}:=\nabla_\mu A_\nu-\nabla_\nu A_\mu$.

\section{Preliminaries}
\subsection{System}
We consider the Einstein-Maxwell system in $n(\ge 3)$ spacetime dimensions with a massless scalar field, which is defined by the action
\begin{align}
\label{action}
S[g_{\mu\nu}, A_{\mu},\phi]=&\int \D^nx\sqrt{-g}\biggl(\frac{1}{2\kappa_{n}}{\cal R}-\frac{1}{4}F_{\mu\nu}F^{\mu\nu}-\frac12({\nabla} \phi)^2 \biggl). 
\end{align}
This action gives the following field equations:
\begin{align}
E_{\mu\nu}:={\cal R}_{\mu\nu}-\frac12g_{\mu\nu}{\cal R}-\kappa_{n}\left(T^{(\rm em)}_{\mu\nu} +T^{(\phi)}_{\mu\nu}\right)=0, \qquad \nabla_\nu F^{\mu\nu}=0, \qquad \dalm\phi=0, \label{em-kg}
\end{align}
where the energy-momentum tensors for the Maxwell field and the massless Klein-Gordon field are
\begin{align}
T^{(\rm em)}_{\mu\nu}:=&F_{\mu\rho}F_\nu^{~\rho}-\frac 14 g_{\mu\nu}F_{\rho\sigma}F^{\rho\sigma}, \label{Tab-Max}\\
T^{(\phi)}_{\mu\nu}: =&(\nabla_\mu \phi)(\nabla_\nu \phi)-\frac12 g_{\mu\nu} (\nabla\phi)^2,\label{Tab-scalar}
\end{align}
respectively.

In the present paper, we consider static spacetimes which correspond to a warped product $M^2 \times K^{n-2}$, where $K^{n-2}$ is a $(n-2)$-dimensional Einstein space. A general metric in such a spacetime can be written as
\begin{align}
\D s^2=&g_{tt}(x)\D t^2+g_{xx}(x)\D x^2+R(x)^2\gamma_{ab}(z)\D z^a\D z^b. \label{static}
\end{align}
Here $\gamma_{ab}(z)$ is the metric on the $(n-2)$-dimensional Einstein space $K^{n-2}$, whose Ricci tensor is given by ${}^{(n-2)}{\cal R}_{ab}=k(n-3)\gamma_{ab}$, where $k =1,0,-1$\footnote{Due to the form of \eqref{static}, just the Ricci tensor of $K^{n-2}$ is necessary to write the field equations. The  decomposition of the curvature of a spacetime of the class \eqref{static} can be found in \cite{Dotti-Gleiser}. }.
In addition, we assume $\phi=\phi(x)$ and $A_\mu \D x^\mu =A_t(x)\D t$.
Then, the Maxwell equation and the Klein-Gordon equation (\ref{em-kg}) are integrated to give
\begin{align}
F_{xt}=q\biggl(-\frac{g_{tt}g_{xx}}{R^{2(n-2)}}\biggl)^{1/2},\qquad \frac{\D \phi}{\D x}=\phi_1\biggl(-\frac{g_{xx}}{g_{tt}R^{2(n-2)}}\biggl)^{1/2},
\end{align}
where $q$ and $\phi_1$ are integration constants.
We assume that $\phi_1$ is real and non-zero, namely we are considering a non-constant real scalar field throughout this paper.

\subsection{Absence of Killing horizon}
In the spacetime (\ref{static}), $g_{tt}(x)=0$ corresponds to a Killing horizon if it is regular and not infinity. 
However, it is shown that there is no Killing horizon in the present system unless $\phi_1=0$.

Here we adopt the following coordinates: 
\begin{align}
\D s^2=&-f(x)e^{-2\delta(x)}\D t^2+f(x)^{-1}\D x^2+R(x)^2\gamma_{ab}(z)\D z^a\D z^b.
\end{align}
If $R$ is constant, the Einstein equations give
\begin{align}
0=E^{t}_{~t}-E^{x}_{~x}=-\frac{\phi_1^2e^{2\delta}}{fR^{2(n-2)}}
\end{align}
from which $\phi_1=0$ is concluded.
Since we consider the case with $\phi_1\ne 0$, $R$ is not constant and then we can choose coordinates such that $R(x)=x(\ge 0)$ without loss of generality.
Then, the trace of the Einstein equations gives 
\begin{align}
(n-2){\cal R}=\kappa_{n}\biggl\{-\frac{(n-4)q^2}{x^{2(n-2)}}+\frac{(n-2)\phi_1^2e^{2\delta}}{fx^{2(n-2)}}\biggl\}. \label{killing}
\end{align}
$f(x_{\rm h})=0$ with $|\delta(x_{\rm h})|<\infty$ defines a Killing horizon if $x=x_{\rm h}(>0)$ is not infinity.
Equation~(\ref{killing}) shows that $\lim_{x\to x_{\rm h}}{\cal R}\to \infty$ unless $\phi_1=0$.
Therefore, it is concluded that there is no Killing horizon (and then no event horizon exists), in the presence of a non-constant scalar field.

\section{General solution in four and higher dimensions}
In this section, we present all the static solutions in the present system in four and higher dimensions.
We perform the complete classification in the following coordinates:
\begin{align}
\D s^2=&-F(x)^{-2}\D t^2+F(x)^{2/(n-3)}G(x)^{-(n-4)/(n-3)}\biggl(\D x^2+G(x)\gamma_{ab}(z)\D z^a\D z^b\biggl),\label{gauge-higher}
\end{align}
in which we have
\begin{align}
F_{xt}=\frac{q}{F^{2}G},\qquad \frac{\D \phi}{\D x}=\frac{\phi_1}{G}.
\end{align}

First we derive the basic equations.
The combination $E^x_{~x}+E^a_{~a}=0$ gives the master equation for $G(x)$:
\begin{align}
\frac{\D ^2G}{\D x^2}-2k(n-3)^2=0. \label{master-G}
\end{align}
Using Eq.~(\ref{master-G}), we write the Einstein equations as
\begin{align}
0=E^t_{~t}=&-\frac{n-2}{8(n-3)}F^{-2/(n-3)}G^{(n-4)/(n-3)}\biggl\{-8F^{-1}\frac{\D ^2F}{\D x^2}+4F^{-2}\biggl(\frac{\D F}{\D x}\biggl)^2 \nonumber \\
&-8F^{-1}G^{-1}\frac{\D F}{\D x}\frac{\D G}{\D x}+G^{-2}\biggl(\frac{\D G}{\D x}\biggl)^2 -4k(n-3)^2G^{-1}\biggl\} \nonumber \\
&-\kappa_{n} F^{-2/(n-3)}G^{(n-4)/(n-3)}\biggl(-\frac{q^2}{2}F^{-2}G^{-2}-\frac{\phi_1^2}{2}G^{-2}\biggl),\label{constraint} \\
0=E^x_{~x}=&-\frac{n-2}{8(n-3)}F^{-2/(n-3)}G^{(n-4)/(n-3)}\biggl\{4F^{-2}\biggl(\frac{\D F}{\D x}\biggl)^2-G^{-2}\biggl(\frac{\D G}{\D x}\biggl)^2 +4k(n-3)^2G^{-1}\biggl\} \nonumber \\
&-\kappa_{n} F^{-2/(n-3)}G^{(n-4)/(n-3)}\biggl(-\frac{q^2}{2}F^{-2}G^{-2}+\frac{\phi_1^2}{2}G^{-2}\biggl),
\end{align}
from which we obtain the master equation for $F(x)$:
\begin{align}
F^{-1}\frac{\D ^2F}{\D x^2}+F^{-1}G^{-1}\frac{\D F}{\D x}\frac{\D G}{\D x}-\frac14G^{-2}\biggl(\frac{\D G}{\D x}\biggl)^2 
 +k(n-3)^2G^{-1}+\frac{(n-3)\kappa_{n} \phi_1^2}{n-2}G^{-2}=0. \label{master-F1}
\end{align}

We are now ready to perform the classification.
$G(x)$ and $F(x)$ are obtained from Eqs.~(\ref{master-G}) and (\ref{master-F1}), respectively, and Eq.~(\ref{constraint}) is a constraint on them.

In order to find the location of the (naked) curvature singularities in this coordinate system, we use the trace of the Einstein equations:
\begin{align}
(n-2){\cal R}=&\kappa_{n}\biggl\{-\frac{(n-4)q^2}{(F^2G)^{(n-2)/(n-3)}}+\frac{(n-2)\phi_1^2}{(F^2G^{n-2})^{1/(n-3)}}\biggl\} \nonumber \\
=&\frac{\kappa_{n}\{-(n-4)q^2+(n-2)\phi_1^2F^2\}}{(F^2G)^{(n-2)/(n-3)}}. \label{Ricciscalar}
\end{align}
The above expression shows that the real zeros of $F^2G=0$ correspond to curvature singularities. Note that where the numerator of Eq.~(\ref{Ricciscalar}) vanishes, $F$ has a finite value. Then, the denominator of (\ref{Ricciscalar}) can be zero at that point only  if $G$ vanishes there. However, for all our solutions in which $G$ can be zero at some point, the function $F$ diverges there. Therefore, the numerator and denominator in (\ref{Ricciscalar}) do not vanish simultaneously.  Consequently, in all the solutions presented in this paper, there appear two classes of curvature singularities: One is given by $G=0$ with infinite $F$ satisfying $F^2G=0$ and the other is given by $F=0$ with finite $G$.
At the singularity in the first class, the scalar field diverges, while it remains finite at the singularity in the second class.

\subsection{General solution for $k=1,-1$ when $G(x)$ has real roots}
First we consider the case of $k=1,-1$ where $G(x)$ has real roots.
In this case, Eq.~(\ref{master-G}) is integrated to give
\begin{align}
G(x)=k(n-3)^2(x-a)(x-b), \label{sol-G1}
\end{align}
where $a,b$ are constants.
We can assume $a\ge b>0$ without loss of generality.
Then the scalar field is given by 
\begin{align}
\phi(x)=\phi_0+\frac{\phi_1}{k(n-3)^2(a-b)}\ln\biggl(\varepsilon\frac{x-a}{x-b}\biggl) \label{sol-phi1}
\end{align}
for $a\ne b$  and 
\begin{align}
\phi(x)=\phi_0-\frac{\phi_1}{k(n-3)^2(x-a)} \label{sol-phi2}
\end{align}
for $a=b$.
The scalar field diverges only at $x=a$ and $x=b$.
We have put  $\varepsilon=\pm 1$ in order to make inside the bracket being positive depending on the domain of $x$.

\subsubsection{Type-I solution}
The general solution for Eq.~(\ref{master-F1}) with the constraint (\ref{constraint}) in the case of $a\ne b$ and $\phi_1^2 < (n-2)(n-3)^3(a-b)^2/(4\kappa_{n})$ (corresponding to $\alpha\ne 0$) is
\begin{align}
F(x)=& A\biggl(\varepsilon\frac{x-a}{x-b}\biggl)^{\alpha/2}+B\biggl(\varepsilon\frac{x-a}{x-b}\biggl)^{-\alpha/2}, \label{sol-F1}
\end{align}
where constants $\alpha$, $A$, and $B$ satisfy
\begin{align}
AB=&-\frac{\kappa_{n} q^2}{(n-2)(n-3)^3(a-b)^2\alpha^2 }, \label{sol-rel1}\\
\phi_1^2 =& \frac{(n-2)(n-3)^3(1-\alpha^2)(a-b)^2}{4\kappa_{n}}. \label{sol-rel2}
\end{align}
By Eq.~(\ref{sol-rel2}), the scalar field in this solution is given by 
\begin{align}
\phi(x)=\phi_0\pm\sqrt{ \frac{(n-2)(1-\alpha^2)}{4\kappa_{n} (n-3)}}\ln\biggl(\varepsilon\frac{x-a}{x-b}\biggl).\label{Type-I-phi}
\end{align}
This is the generalization in arbitrary dimensions of the Penney solution  ($n=4$ and $k=1$) found in~\cite{Penney1969}.

In order to identify the location of the singularities, we compute
\begin{align}
F^2G=&k(n-3)^2\biggl\{ A\frac{(x-a)^{(\alpha+1)/2}}{(x-b)^{(\alpha-1)/2}}+B\frac{(x-a)^{-(\alpha-1)/2}}{(x-b)^{-(\alpha+1)/2}}\biggl\}^2,
\end{align}
where we have set $\varepsilon=1$ for simplicity.
Since reality of the scalar field requires $-1<\alpha<1$ by Eq.~(\ref{sol-rel2}), $F^2G=0$ holds at both $x=a$ and $x=b$ and hence they are curvature singularities.
A solution of $F(x)=0$ for $AB< 0$ necessarily satisfies $x\ne a,b$ and it corresponds to a curvature singularity with finite $\phi$.

The neutral limit $q\to 0$ of this type-I solution is realized for $AB\alpha (a-b)\to 0$, where the scalar field remains nontrivial only for $AB\alpha \to 0$.
The limit $A=0$ or $B=0$ gives the higher-dimensional and topological generalization of the Janis-Newman-Winicour solution~\cite{jnw1968,Fisher1948,wyman1981,JNWhigher,AbdolrahimiShoom2010} (the generalized JNW solution, hereafter), of which friendly form is given by 
\begin{align}
\D s^2=&-f(r)^\alpha \D { t}^2+f(r)^{-\alpha/(n-3)}\biggl(f(r)^{-(n-4)/(n-3)}\D r^2+r^2f(r)^{1/(n-3)}\gamma_{ab}(z)\D z^a\D z^b\biggl),\label{JNW-higher}\\
\phi =&\phi_0\pm\sqrt{\frac{(n-2)(1-\alpha^2)}{4\kappa_ {n}(n-3)}}\ln f(r), \qquad f(r)=k-\frac{\mu}{r^{n-3}},
\end{align}
where $\mu$ is a constant and $t$ has been rescaled by a constant.
The limit $\alpha=0$ of the type-I solution also gives the above solution with $\alpha=0$.
The limit $a=b$ gives a warped-product spacetime of a two-dimensional Minkowski spacetime and a $(n-2)$-dimensional Ricci-flat space, which we call the Ricci-flat-A solution hereafter.
The simplest form of the Ricci-flat-A solution is given by 
\begin{align}
\D s^2=-\D { t}^2+k^{-1}\D r^2+r^2\gamma_{ab}(z)\D z^a\D z^b, \label{Ricci-flat}
\end{align}
which is static only for $k=1$.
In contrast, $k=-1$ gives a dynamical spacetime, which is shown by the coordinate transformations $T=r$ and $X=i{t}$.

The trivial scalar-field limit $\phi\to $ constant of the type-I solution is realized for $(1-\alpha^2)(a-b)\to 0$.
The limit $a=b$ gives the Ricci-flat-A solution \eqref{Ricci-flat}, while the limit $\alpha=\pm 1$ gives the higher-dimensional and topological generalization of Reissner-Nordstr\"om solution (the generalized Reissner-Nordstr\"om solution, hereafter), of which simple form is given by 
\begin{align}
\D s^2=&-f(r)\D { t}^2+f(r)^{-1}\D r^2+r^2\gamma_{ab}(z)\D z^a\D z^b,\label{RN-higher}\\
F_{rt} =&\sqrt{\frac{(n-2)(n-3)}{\kappa_{n}}}\frac{Q}{r^{n-2}}, \qquad f(r)=k-\frac{\mu}{r^{n-3}}+\frac{Q^2}{r^{2(n-3)}},
\end{align}
where $\mu$ and $Q$ are constants related with the mass and electric charge of the solution, respectively.

\subsubsection{Type-II solution}

The general solution for Eq.~(\ref{master-F1}) with the constraint (\ref{constraint}) in the case of $a\ne b$ and $\phi_1^2 > (n-2)(n-3)^3(a-b)^2/(4\kappa_n)$ (corresponding to ${\bar \alpha}\ne 0$) is
\begin{align}
F(x)=& F_1\cos\biggl\{\frac{{\bar \alpha}}{2}\ln
\biggl(\varepsilon\frac{x-a}{x-b}\biggl)\biggl\}+
F_2\sin\biggl\{\frac{{\bar \alpha}}{2}\ln
\biggl(\varepsilon\frac{x-a}{x-b}\biggl)\biggl\}, \label{type-I-g1000},
\end{align}
where constants ${\bar \alpha}$, $F_1$, and $F_2$ satisfy
\begin{align}
F_1^2+F_2^2=&\frac{4\kappa_n q^2}{(n-2)(n-3)^3(a-b)^2{\bar \alpha}^2}, \label{sol-rel1b}\\
\phi_1^2 =& \frac{(n-2)(n-3)^3(1+{\bar \alpha}^2)(a-b)^2}{4\kappa_n}. \label{sol-rel2b}
\end{align}
The scalar field is given by 
\begin{align}
\phi(x)=&\phi_0\pm\sqrt{ \frac{(n-2)(1+{\bar \alpha}^2)}{4\kappa_n (n-3)}}\ln\biggl(\varepsilon\frac{x-a}{x-b}\biggl).\label{Type-I-phib}
\end{align}

This solution corresponds to the type-I solution with a pure imaginary $\alpha$.
Equation~(\ref{Type-I-phi}) shows that the scalar field remains real for a pure imaginary value of $\alpha$ satisfying $-1<\alpha^2<0$.
Indeed, the above type-II solution is obtained from the following alternative expression of the metric function (\ref{sol-F1}) of the type-I solution;
\begin{align}
F(x)=& (A+B)\cosh\biggl\{\frac{\alpha}{2}\ln\biggl(\varepsilon\frac{x-a}{x-b}\biggl)\biggl\}+(A-B)\sinh\biggl\{\frac{\alpha}{2}\ln\biggl(\varepsilon\frac{x-a}{x-b}\biggl)\biggl\} \label{type-I-real1000}
\end{align}
with a pure imaginary $\alpha$.
The relations between the constants in the type-I and type-II solutions are 
\begin{align}
F_1=A+B,\qquad F_2=i(A-B), \qquad \alpha=i{\bar \alpha}.
\end{align}

Since the metric function $F$ (\ref{type-I-g1000}) is finite in this type-II solution, the zeros of $G$, namely $x=a$ and $x=b$, correspond to curvature singularities.
Moreover, there is an infinite number of solutions for $F(x)=0$, which all correspond to curvature singularities
are all different from $x=a$ or $x=b$ and represent curvature singularities with finite $\phi$.

In this solution, a neutral limit is given by $(a-b){\bar \alpha}=0$.
The limit ${\bar\alpha}=0$ gives the generalized JNW solution (\ref{JNW-higher}) with $\alpha=0$, while the limit $a=b$ gives the Ricci-flat-A solution (\ref{Ricci-flat}).
On the other hand, the trivial scalar-field limit $\phi\to $ constant is realized for $a=b$, which gives the Ricci-flat-A solution \eqref{Ricci-flat}.

\subsubsection{Type-III solution}
In the case of $a\ne b$ and $\phi_1^2=  (n-2)(n-3)^3(a-b)^2/(4\kappa_n)$, the general solution is
\begin{align}
F(x)=A\ln\biggl(\varepsilon\frac{x-a}{x-b}\biggl)+B,\label{sol-F2}
\end{align}
where
\begin{align}
A^2=\frac{\kappa_{n} q^2}{ (n-2)(n-3)^3(a-b)^2}. \label{sol-rel3}
\end{align}
The scalar field in this solution is given by 
\begin{align}
\phi(x)=\phi_0\pm\sqrt{ \frac{n-2}{4\kappa_n (n-3)}}\ln\biggl(\varepsilon\frac{x-a}{x-b}\biggl).
\end{align}

In this solution, both $x=a$ and $x=b$ correspond to curvature singularities because the following expression
\begin{align}
F^2G=&k(n-3)^2(x-a)(x-b)\biggl\{A\ln\biggl(\varepsilon\frac{x-a}{x-b}\biggl)+B\biggl\}^2
\end{align}
shows that $F^2G=0$ holds there.
A solution of $F(x)=0$ for $A\ne  0$ satisfies $x\ne a,b$ and it corresponds to a curvature singularity with finite $\phi$.

The neutral limit $q\to 0$ and the trivial scalar-field limit of this type-III solution are realized for $A(a-b)\to 0$ and $a-b\to 0$, respectively.
In the limit of $A=0$, the solution reduces to the generalized JNW solution (\ref{JNW-higher}) with $\alpha=0$, while the solution becomes the Ricci-flat-A solution \eqref{Ricci-flat} in the limit of $b\to a$.

\subsubsection{Type-IV solution}
Lastly, in the case of $a=b$, the general solution is 
\begin{align}
F(x)=&A\sin\biggl(\sqrt{\frac{\kappa_{n}}{(n-2)(n-3)}}\frac{\phi_1}{k(n-3)(x-a)}\biggl)
\nonumber \\
+&B\cos\biggl(\sqrt{\frac{\kappa_{n}}{(n-2)(n-3)}}\frac{\phi_1}{k(n-3)(x-a)}\biggl),\label{sol-F3}
\end{align}
where
\begin{align}
q^2 = (A^2+B^2)\phi_1^2. \label{sol-rel4}
\end{align}
The scalar field in this solution is given by Eq.~(\ref{sol-phi2}).
Since the metric function $F$ is finite in this solution, the zero of $G$, $x=a$, corresponds to a curvature singularity.
Moreover, in this special case, there is an infinite number of solutions for $F(x)=0$, which are all different from $x=a$ and represent curvature singularities with finite $\phi$.

The neutral limit $q\to 0$ and the trivial scalar-field limit of this type-IV solution are equivalent, in which the solution becomes the Ricci-flat-A solution given by \eqref{Ricci-flat}.

\subsection{General solution for $k=1,-1$ when $G(x)$ has no real root}
\label{sec:noroot}
Next we consider the case of $k=1,-1$ where $G(x)$ has no a real root.
In this case, Eq.~(\ref{master-G}) is integrated to give
\begin{align}
G(x)=k(n-3)^2x^2+G_0, \label{sol-G2}
\end{align}
where we have used the degree of freedom to change the origin of $x$ and $k G_0$ must be positive for ensuring the absence of real roots. However, $G(x)$ must be positive for keeping the signature of the metric (and its reality in some dimensions). Then,  $k=1$ and $G_0>0$ is the unique option in this case.

From \eqref{sol-G2}, the scalar field is given by 
\begin{align}
\phi(x)=\phi_0+\frac{\phi_1}{(n-3)\sqrt{kG_0}}\arctan\biggl(\frac{(n-3)kx}{\sqrt{kG_0}}\biggl). \label{sol-phi3}
\end{align}
Remarkably, the scalar field is finite everywhere in this class of solutions.

\subsubsection{Type-V${}_1$ solution}
The general solution for Eq.~(\ref{master-F1}) with the constraint (\ref{constraint}) is given by 
\begin{align}
F(x)=&A\sin\biggl\{\sqrt{\frac{\kappa_{n}\phi_1^2+(n-2)(n-3)kG_0}{(n-2)(n-3)kG_0}}\arctan\biggl(\frac{(n-3)kx}{\sqrt{kG_0}}\biggl)\biggl\} \nonumber \\
+&B\cos\biggl\{\sqrt{\frac{\kappa_{n}\phi_1^2+(n-2)(n-3)kG_0}{(n-2)(n-3)kG_0}}\arctan\biggl(\frac{(n-3)kx}{\sqrt{kG_0}}\biggl)\biggl\},\label{sol-F4}
\end{align}
where constants $A$ and $B$ satisfy
\begin{align}
A^2+B^2=\frac{\kappa_{n} q^2}{(n-2)(n-3)kG_0+\kappa_{n}\phi_1^2}. \label{sol-rel5}
\end{align}
The scalar field in this solution is given by Eq.~(\ref{sol-phi3}).
In this case, there is an infinite number of solutions for $F(x)=0$, which all correspond to curvature singularities with finite $\phi$.

There is no neutral limit $q\to 0$ of this type-V${}_1$ solution under the assumption that the scalar field is real.
On the other hand, the trivial scalar-field limit $\phi\to $ constant is realized for $\phi_1\to 0$, in which we have
\begin{align}
F(x)=&\frac{A(n-3)kx+B\sqrt{kG_0}}{\sqrt{kG(x)}}.
\end{align}
The solution in this limit is the generalized Reissner-Nordstr\"om solution (\ref{RN-higher}) for $A\ne 0$.
For $A=0$, the limit is the Bertotti-Robinson-type cross-product solution, of which a readily form is given by 
 \begin{align}
\D s^2=&-\biggl(1+\frac{k(n-3)^2r^2}{r_0^2}\biggl)\D { t}^2+\biggl(1+\frac{k(n-3)^2r^2}{r_0^2}\biggl)^{-1}\D r^2+r_0^2\gamma_{ab}(z)\D z^a\D z^b,\label{BR-higher}\\
&F_{rt} =\sqrt{\frac{(n-2)(n-3)}{\kappa_{n}}}\frac{Q}{r_0^{n-2}}, \qquad Q^2=kr_0^{2(n-3)},
\end{align}
where $t$ has been rescaled and $r_0^2=G_0$.

\subsection{General solution for $k=0$}
In the case of $k=0$, Eq.~(\ref{master-G}) is integrated to give
\begin{align}
G(x)=G_1x+G_0, \label{sol-G3}
\end{align}
where $G_0$ and $G_1$ are constants.
Then the scalar field is given by 
\begin{align}
\phi(x)=\phi_0+\frac{\phi_1}{G_1}\ln\biggl\{\varepsilon(G_1x+G_0)\biggl\}  \label{sol-phi4}
\end{align}
for $G_1\ne 0$ and 
\begin{align}
\phi(x)=\phi_0+\frac{\phi_1}{G_0}x \label{sol-phi5}
\end{align}
for $G_1=0$.

\subsubsection{Type-VI${}_0$ solution}
The general solution for Eq.~(\ref{master-F1}) with the constraint (\ref{constraint}) in the case of  $G_1\ne 0$ and $\phi_1^2< (n-2)G_1^2/(4(n-3)\kappa_{n})$  (corresponding to $\alpha\ne 0$) is given by
\begin{align}
F(x)=&A\biggl\{\varepsilon(G_1x+G_0)\biggl\}^{\alpha/2}+B\biggl\{\varepsilon(G_1x+G_0)\biggl\}^{-\alpha/2},\label{sol-F6}
\end{align}
where constants $\alpha$, $A$, and $B$ satisfy
\begin{align}
\phi_1^2=&\frac{(n-2)(1-\alpha^2)G_1^2}{4(n-3)\kappa_{n}}, \label{sol-real-cond1} \\
AB=&-\frac{(n-3)\kappa_{n} q^2}{(n-2)\alpha^2G_1^2}. \label{sol-rel6}
\end{align}
By Eq.~(\ref{sol-real-cond1}), the scalar field in this solution is given by 
\begin{align}
\phi(x)=\phi_0\pm\sqrt{\frac{(n-2)(1-\alpha^2)}{4(n-3)\kappa_{n}}}\ln\biggl\{\varepsilon(G_1x+G_0)\biggl\}.\label{Type-V-phi}
\end{align}

In this case, we compute
\begin{align}
F^2G=&\biggl\{A(G_1x+G_0)^{(1+\alpha)/2}+B(G_1x+G_0)^{(1-\alpha)/2}\biggl\}^2,
\end{align}
where we have set $\varepsilon=1$ for simplicity.
Since reality of the scalar field requires $-1<\alpha<1$ by Eq.~(\ref{sol-real-cond1}), $F^2G=0$ holds at $G=0$, namely $x=-G_0/G_1$. 
Hence it corresponds to a curvature singularity.
Also, a solution of $F(x)=0$ for $AB< 0$ corresponds to a curvature singularity but with finite $\phi$.

The neutral limit $q\to 0$ of this type-VI${}_0$ solution is realized for $AB\alpha G_1\to 0$, where the scalar field remains nontrivial only for $AB\alpha \to 0$.
The limit $A=0$ or $B=0$ gives the generalized JNW solution (\ref{JNW-higher}) with $k=0$.
The limit $\alpha=0$ of this solution also gives the generalized JNW solution but with $\alpha=0$.
The limit $G_1=0$ gives a direct-product spacetime of a two-dimensional Minkowski spacetime and a $(n-2)$-dimensional Ricci-flat space, which we call the Ricci-flat-B solution hereafter.
The simplest form of the Ricci-flat-B solution is given by 
\begin{align}
\D s^2=-\D { t}^2+\D r^2+\gamma_{ab}(z)\D z^a\D z^b, \label{Ricci-flat-B}
\end{align}
which is different from the Ricci-flat-A solution \eqref{Ricci-flat}.

The trivial scalar-field limit $\phi\to $ constant of the type-VI${}_0$ solution is realized for $(1-\alpha^2)G_1\to 0$.
The limit $G_1=0$ gives the Ricci-flat-B solution \eqref{Ricci-flat-B}, while the limit $\alpha=\pm 1$ gives the generalized Reissner-Nordstr\"om solution (\ref{RN-higher}) with $k=0$.

\subsubsection{Type-VII${}_0$ solution}
 
The general solution for Eq.~(\ref{master-F1}) with the constraint (\ref{constraint}) in the case of  $G_1\ne 0$ and $\phi_1^2> (n-2)G_1^2/(4(n-3)\kappa_{n})$ (corresponding to ${\bar \alpha}\ne 0$) is given by 
\begin{align}
F(x)=&F_1\cos\biggl\{\frac{{\bar\alpha}}{2}\ln\biggl(\varepsilon(G_1x+G_0)\biggl)\biggl\}+F_2\sin\biggl\{\frac{{\bar\alpha}}{2}\ln\biggl(\varepsilon(G_1x+G_0)\biggl)\biggl\},\label{type-Vb-g1000}
\end{align}
where constants ${\bar \alpha}$, $F_1$, and $F_2$ satisfy
\begin{align}
&\phi_1^2=\frac{(n-2)(1+{\bar\alpha}^2)G_1^2}{4(n-3)\kappa_{n}}, \label{sol-rel1b7}\\
&F_1^2+F_2^2=\frac{4(n-3)\kappa_{n} q^2}{(n-2){\bar\alpha}^2G_1^2}. \label{sol-rel2b7}
\end{align}
The scalar field is given by 
\begin{align}
\phi(x)=&\phi_0\pm\sqrt{\frac{(n-2)(1+{\bar\alpha}^2)}{4(n-3)\kappa_{n}}}\ln\biggl\{\varepsilon(G_1x+G_0)\biggl\}. \label{type-Vb-phi1000}
\end{align}

This solution actually corresponds to the type-VI${}_0$ solution with a pure imaginary $\alpha$.
Equation~(\ref{Type-V-phi}) shows that the scalar field remains real for a pure imaginary value of $\alpha$ satisfying $-1<\alpha^2<0$.
Indeed, the above solution is obtained from the following alternative expression of the metric function (\ref{sol-F6}) of the type-VI${}_0$ solution
\begin{align}
F(x)=&(A+B)\cosh\biggl\{\frac{\alpha}{2}\ln\biggl(\varepsilon(G_1x+G_0)\biggl)\biggl\}+(A-B)\sinh\biggl\{\frac{\alpha}{2}\ln\biggl(\varepsilon(G_1x+G_0)\biggl)\biggl\}. \label{type-V-real1000}
\end{align}
with a pure imaginary $\alpha$.
The relations between the constants in the type-VI${}_0$ and type-VII${}_0$ solutions are 
\begin{align}
F_1=A+B,\qquad F_2=i(A-B), \qquad \alpha=i{\bar \alpha}.
\end{align}

Since the metric function $F$ (\ref{type-Vb-g1000}) is finite in this type-VII${}_0$ solution, the zero of $G$, $x=-G_0/G_1$, corresponds to a curvature singularity.
Moreover, there is an infinite number of solutions for $F(x)=0$, which all correspond to curvature singularities
are all different from $x=-G_0/G_1$ and represent curvature singularities with finite $\phi$.

The neutral limit $q\to 0$ of this type-VII${}_0$ solution is realized for ${\bar\alpha} G_1\to 0$, where the scalar field remains nontrivial only for ${\bar \alpha} \to 0$.
The limit ${\bar \alpha}=0$ of this solution gives the generalized JNW solution with $\alpha=0$.
The limit $G_1=0$ gives the Ricci-flat-B solution \eqref{Ricci-flat-B}.
The trivial scalar-field limit $\phi\to $ constant is realized only for $G_1\to 0$, which gives the Ricci-flat-B solution \eqref{Ricci-flat-B}.

\subsubsection{Type-VIII${}_0$ solution}
The general solution for Eq.~(\ref{master-F1}) in the case of $G_1\ne 0$ and $\phi_1^2=(n-2)G_1^2/(4(n-3)\kappa_{n})$ is given by 
\begin{align}
F(x)=&A\ln\biggl\{\varepsilon(G_1x+G_0)\biggl\}+B,\label{sol-F7}
\end{align}
where
\begin{align}
A^2 = \frac{(n-3)\kappa_{n} q^2}{(n-2) G_1^2}. \label{sol-rel7}
\end{align}
The scalar field in this solution is given by 
\begin{align}
\phi(x)=\phi_0\pm\sqrt{\frac{n-2}{4(n-3)\kappa_{n}}}\ln\biggl\{\varepsilon(G_1x+G_0)\biggl\}. \label{sol-phi-VI2000}
\end{align}

Also in this case, $G(x)=0$ corresponds to a curvature singularity since $F^2G=0$ holds there.
In addition, a solution of $F(x)=0$ for $A\ne 0$ corresponds to a curvature singularity with finite $\phi$.

The neutral limit $q\to 0$ and the trivial scalar-field limit of this type-VIII${}_0$ solution are realized for $AG_1\to 0$ and $G_1\to 0$, respectively.
In the limit of $A=0$, the solution reduces to the generalized JNW solution (\ref{JNW-higher}) with $k=0$ and $\alpha=0$, while the solution becomes the Ricci-flat-B solution \eqref{Ricci-flat-B} in the limit of $G_1\to 0$.

\subsubsection{Type-IX${}_0$ solution}
Lastly, the general solution in the case of $G_1= 0$ is given by 
\begin{align}
F(x)=&A\sin\biggl(\sqrt{\frac{(n-3)\kappa_{n}}{n-2}}\frac{\phi_1}{G_0}x\biggl)+B\cos\biggl(\sqrt{\frac{(n-3)\kappa_{n}}{n-2}}\frac{\phi_1}{G_0}x\biggl),\label{sol-F8}
\end{align}
where
\begin{align}
A^2+B^2=\frac{q^2}{\phi_1^2 }. \label{sol-rel8}
\end{align}
The scalar field in this solution is given by Eq.~(\ref{sol-phi5}).
Since $F$ is finite everywhere, $G(x)=0$ corresponds to a curvature singularity.
Additionally, in this special case, there is an infinite number of zeros of $F(x)=0$ which are all curvature singularities with finite $\phi$.

The neutral limit $q\to 0$ and the trivial scalar-field limit of this type-IX${}_0$ solution are equivalent, in which the solution reduces to the Ricci-flat-B solution \eqref{Ricci-flat-B}.

\section{General solution in three dimensions}
In this section, we present the classification in three dimensions.
We will use the Einstein equations in the form of ${\cal E}^\mu_{~~\nu}=0$, where
\begin{align}
{\cal E}_{\mu\nu}:={\cal R}_{\mu\nu}-\kappa_{n}\biggl\{F_{\mu\rho}F_\nu^{~\rho}-\frac{1}{2(n-2)}g_{\mu\nu}F_{\rho\sigma}F^{\rho\sigma}\biggl\} -\kappa_{n}(\nabla_\mu \phi )(\nabla_\nu \phi). \label{EFE-different}
\end{align}
Clearly the gauge (\ref{gauge-higher}) does not work for $n=3$.
For the three-dimensional case, we adopt the following coordinates:
\begin{align}
\D s^2=-e^{-2\Phi(r)}\D t^2+e^{2\Psi(r)}(\D r^2+e^{2\Phi(r)}\D\theta^2).
\end{align}
In this coordinate system, the scalar field is integrated to give
\begin{align}
\phi(r)=\phi_0+\phi_1 r,
\end{align}
where $\phi_0$ and $\phi_1$ are constants. Moreover,  the field strength is given by
\begin{align}
F_{rt}=q e^{-2\Phi(r)},
\end{align}
where $q$ is an integration constant.

Now the Einstein equations are written as
\begin{align}
&\frac{\D^2\Phi}{\D r^2}=0,\label{3dim-1}\\
&2\frac{\D \Phi}{\D r}\frac{\D \Psi}{\D r}+2\biggl(\frac{\D \Phi}{\D r}\biggl)^2+\frac{\D^2 \Psi}{\D r^2}=-\kappa_{3} \phi_1^2,\label{3dim-2}\\
&\frac{\D^2 \Psi}{\D r^2}=-\kappa_{3} q^2e^{-2\Phi}.\label{3dim-3}
\end{align}
The general solution of Eq.~(\ref{3dim-1}) is given by 
\begin{equation}
e^{-2 \Phi}=c_0^2 e^{ -2 \Phi_1 r},
\end{equation}
where $c_0$ and $\Phi_1$ are constants.
The classification is rather simple: We solve Eq.~(\ref{3dim-3}) for $\Psi(r)$ and use Eq.~(\ref{3dim-2}) as a constraint.

\subsection{General solution for $\Phi_1= 0$: Type-X${}_3$ and XI${}_3$ solutions}
If $\Phi_1=0$, the general solution for $\Psi(r)$ is
\begin{align}
\Psi(r)=-\frac12\kappa_3 \phi_1^2(r-a)(r-b)
\end{align}
if $\Psi(r)$ has real roots and 
\begin{align}
\Psi(r)=-\Psi_0-\frac12\kappa_3 \phi_1^2r^2
\end{align}
if $\Psi(r)$ has no real root, where $a$, $b$, and $\Psi_0(>0)$ are constants.
In both cases, $\phi_1$ is given by  
\begin{align}
\phi_1^2=c_0^2q^2. \label{sol-rel9}
\end{align}
These  solutions acquire a simple form  after a rescaling of $t$ and $\theta$;
\begin{align}
\D s^2=&-\D t^2+e^{-\kappa_{3} \phi_1^2(r-a)(r-b)}(\D r^2+\D\theta^2) \qquad \mbox{[Type-X${}_3$ solution]}\label{sol-metric1}
\end{align}
and 
\begin{align}
\D s^2=&-\D t^2+e^{-2\Psi_0-\kappa_{3} \phi_1^2r^2}(\D r^2+\D\theta^2)  \qquad \mbox{[Type-XI${}_3$ solution]}.\label{sol-metric2}
\end{align}
In both cases, $\phi(r)$ and $F_{rt}$ are given by 
\begin{align}
\phi(r)=&\phi_0\pm\sqrt{q} r,\quad F_{rt}=q. \label{sol-matter1}
\end{align}
In these type-X${}_3$ and XI${}_3$ solutions, the neutral limit $q\to 0$ and the trivial scalar-field limit are equivalent, in which the solution reduces to Minkowski.

Since the Kretschmann invariant $K:={\cal R}_{\mu\nu\rho\sigma}{\cal R}^{\mu\nu\rho\sigma}$ is given by 
\begin{align}
K=4\kappa_{3}^2\phi_1^4e^{2\kappa_3\phi_1^2(r-a)(r-b)}
\end{align}
for the metric (\ref{sol-metric1}) and 
\begin{align}
K=4\kappa_{3}^2\phi_1^4e^{4\Psi_0+2\kappa_3\phi_1^2r^2}
\end{align}
for the metric (\ref{sol-metric2}), curvature singularities are located at $r\to \pm \infty$ in both cases.

\subsection{General solution for $\Phi_1\ne 0$: Type-XII${}_3$ solution}
If $\Phi_1\ne 0$, the general solution for $\Psi(r)$ is
\begin{align}
\Psi(r)=\Psi_0-\biggl(\Phi_1+\frac{\kappa_{3} \phi_1^2}{2\Phi_1}\biggl)r-\frac{\kappa_{3} q^2c_0^2}{4\Phi_1^2 } e^{ -2 \Phi_1 r}.
\end{align}
After the coordinate transformations $x=c_0\Phi_1^{-1}e^{ -\Phi_1 r}$ and $\Phi_1t\to t$, we obtain the solution in the simplest form:
\begin{align}
\D s^2=&- x^2\D t^2+x^{\kappa_{3} \phi_1^2}\exp\biggl(2\Psi_0-\frac12 \kappa_{3} q^2x^2\biggl)(\D x^2+\D \theta^2), \label{sol-metric3}\\
\phi(r)=&\phi_0+\phi_1\ln |x|,\quad F_{xt}=qx, \label{sol-matter2}
\end{align}
where $\phi_0$, $\phi_1$, and $\Psi_0$ have been redefined.

The Kretschmann invariant $K:={\cal R}_{\mu\nu\rho\sigma}{\cal R}^{\mu\nu\rho\sigma}$ is given by 
\begin{align}
K=\kappa_{3}^2(3q^4x^4-2q^2\phi_1^2x^2+3\phi_1^4)x^{-2\kappa_3\phi_1^2-4}e^{\kappa_{3} q^2x^2-4\Psi_0}.
\end{align}
Hence, curvature singularities are located at $x=0,\pm \infty$.
In the neutral limit $q\to 0$, this type-XII${}_3$ solution reduces to the one obtained in~\cite{bbl1986,Virbhadra1995}.
On the other hand, the solution in the trivial scalar-field limit $\phi_1\to 0$ was obtained by several authors independently~\cite{dm1985,gsa1986}.

\section{Non-uniqueness of asymptotically flat solutions}
In the previous two sections, we have obtained the general static solution in the present system in arbitrary $n(\ge 3)$ dimensions.
The general solution consists of nine solutions for $n\ge 4$ and three solutions for $n=3$, which are summarized in Table~\ref{Table:solutions}. 
\begin{table}[htb]
\begin{center}
  \begin{tabular}{|l|c|c|c|c|c|} \hline
    Name& Metric functions & $\phi$  & Phantom  & Comment \\ 
    &  &  & allowed? & \\
    \hline \hline
    Type-I ($k=\pm 1$)& (\ref{sol-G1}), (\ref{sol-F1}) & (\ref{Type-I-phi})  & Yes & $n=4$, $k=1$ given in~\cite{Penney1969} \\ 
   Type-II ($k=\pm 1$)& (\ref{sol-G1}), (\ref{type-I-g1000}) & (\ref{Type-I-phib})  & No &  \\ 
    Type-III ($k=\pm 1$)& (\ref{sol-G1}), (\ref{sol-F2}) & (\ref{sol-phi1})   & No &  $\phi\to$ constant not allowed\\ 
    Type-IV ($k=\pm 1$)& (\ref{sol-G1}), (\ref{sol-F3}) & (\ref{sol-phi2})   & Yes & \\ 
    Type-V${}_1$ ($k= 1$)& (\ref{sol-G2}), (\ref{sol-F4}) & (\ref{sol-phi3})   &  Yes & $q\to 0$ not allowed \\  
   Type-VI${}_0$ ($k=0$)& (\ref{sol-G3}), (\ref{sol-F6}) & (\ref{Type-V-phi})   & Yes & \\ 
   Type-VII${}_0$ ($k=0$)& (\ref{sol-G3}), (\ref{type-Vb-g1000}) & (\ref{type-Vb-phi1000})   & No & \\ 
    Type-VIII${}_0$ ($k=0$)& (\ref{sol-G3}), (\ref{sol-F7}) & (\ref{sol-phi-VI2000})   & No & $\phi\to$ constant not allowed \\ 
    Type-IX${}_0$ ($k=0$)& (\ref{sol-G3}), (\ref{sol-F8}) & (\ref{sol-phi5})   & Yes & \\ 
    Type-X${}_3$ ($n=3$)& (\ref{sol-metric1}) & (\ref{sol-matter1})   & No & \\ 
   Type-XI${}_3$ ($n=3$)& (\ref{sol-metric2}) & (\ref{sol-matter1})   & No & \\ 
    Type-XII${}_3$ ($n=3$)& (\ref{sol-metric3}) & (\ref{sol-matter2})  & Yes & $q=0$~\cite{bbl1986,Virbhadra1995}, $\phi_1=0$~\cite{dm1985,gsa1986}.\\ \hline
  \end{tabular}
  \caption{Classification of the static solutions. The limit $q\to 0$ is allowed in the solutions (except for type-V${}_1$), where $\phi(x)$ then necessarily becomes constant in the solutions IV and IX${}_3$--XI${}_3$.
The limit to constant $\phi$ is allowed in the solutions (except for type-III and VIII${}_0$), where $q$ necessarily reduces to zero in the solutions II, IV, VII${}_0$, IX${}_0$, XI${}_3$, and X${}_3$. The term ``phantom'' in the table means a configuration where the scalar field $\phi$ is pure imaginary.
}
\label{Table:solutions}
\end{center}
\end{table}

Clarifying the limiting cases, we have shown that, in the absence of a Maxwell field ($q=0$), the general static solution is unique; the generalized JNW solution (\ref{JNW-higher}) for $n\ge 4$ and the Virbhadra solution (the type-XII${}_3$ solution (\ref{sol-metric3}) with $q=0$) for $n=3$.
In contrast, in the presence of a nontrivial Maxwell field, the general static solution consists of multiple distinct solutions and therefore the static solution is no longer unique.
Then a natural question arises: Is there a unique static and asymptotically flat solution?

In three dimensions, all of the type X${}_3$--XII${}_3$ solutions are not asymptotically flat for $q\ne 0$ because all the components of the Riemann tensor cannot be zero simultaneously.
However in the neutral case ($q=0$), the unique type-XII${}_3$ (Virbhadra) solution (\ref{sol-metric3}) is asymptotically locally flat for $r\to \infty$.

How about the cases in four and higher dimensions?
In the absence of a massless scalar field, namely in the Einstein-Maxwell system with $n\ge 4$, the general spherically symmetric solution consists of the arbitrary-dimensional Reissner-Nordstr\"om solution (\ref{RN-higher}) and the Bertotti-Robinson solution (\ref{BR-higher}) where the base manifold $K^{n-2}$ is a $(n-2)$-sphere $S^{n-2}$ which gives $k=1$.
Among these two, the former is the unique asymptotically flat solution.
In contrast, if both configurations of the Maxwell field and scalar field are nontrivial, the static and asymptotically flat solution is no longer unique.
In this section, we will show that the general static solution for $n\ge 4$ contains multiple distinct asymptotically flat solutions.

\subsection{Asymptotics}

In this section we consider the $(n-2)-$sphere $S^{n-2}$ as the base manifold $K^{n-2}$, so that $k=1$ and $\gamma_{ab} \D z^a \D z^b=\D\Omega^2_{n-2}$, where $\D\Omega^2_{n-2}$ is the line element on $S^{n-2}$. For $x \to \infty$, the functions $G(x)$ and $F(x)$ appearing in the line element of solutions I--IV and V${}_1$ behave as 
\begin{equation} \label{B}
G(x)=\left\{ \begin{array}{ll}
\displaystyle (n-3)^2\{x^2-(a+b)x+ab\}  & \mbox{[Type-I, II, III, IV]},\\[4mm]
\displaystyle (n-3)^2x^2+G_0   & 
       \mbox{[Type-V${}_1$]}
        .\end{array} \right.
\end{equation}
and 
\begin{equation} \label{F-asymp}
F(x)\simeq \left\{ \begin{array}{ll}
\displaystyle (A+B)-\frac{\alpha(A-B)(a-b)}{2x} & \\[3mm] \displaystyle \qquad\quad  +\frac{\alpha  (a-b) \{\alpha  (a-b) (A+B)-2 (a+b) (A-B)\}}{8 x^2}  & \mbox{[Type-I]},\\[4mm]
\displaystyle F_1-\frac{{\bar \alpha}F_2(a-b)}{2x}-\frac{{\bar \alpha}  (a-b) \{{\bar \alpha}F_1(a-b)+2F_2(a+b)\}}{8 x^2}  & \mbox{[Type-II]},\\[4mm]
\displaystyle B-\frac{A(a-b)}{x}  -\frac{A \left(a^2-b^2\right)}{2 x^2}& 
       \mbox{Type-III]},\\[4mm]
\displaystyle B+\sqrt{\frac{\kappa_{n}}{(n-2)(n-3)^3}}\frac{A\phi_1}{x} & \\[4mm] \displaystyle   \quad +\sqrt{\frac{\kappa_{n}}{(n-2)(n-3)^3}}\frac{\phi_1}{x^2}\left(a A-\sqrt{\frac{\kappa_{n}}{(n-2)(n-3)^3}}\frac{B\phi_1}{2} \right) & 
       \mbox{[Type-IV]},\\[4mm]
       \displaystyle \biggl(A\sin\frac{\eta\pi}{2}+B\cos\frac{\eta\pi}{2}\biggl)-\frac{\sqrt{G_0}\eta}{(n-3)x}\biggl(A\cos\frac{\eta\pi}{2}-B\sin\frac{\eta\pi}{2}\biggl)& \\[4mm] \displaystyle  \qquad   \qquad   \qquad   \qquad  \quad   -\frac{ G_0 \eta^2}{2(n-3)^2 x^2}  \left(A\sin\frac{\eta\pi}{2}+B\cos\frac{\eta\pi}{2}\right)&
       \mbox{[Type-V${}_1$]},
        \end{array} \right.
\end{equation}
where
\begin{align}
\eta:=\sqrt{\frac{\kappa_{n}\phi_1^2+(n-2)(n-3)G_0}{(n-2)(n-3)G_0}}.
\end{align}

All the above cases have the following form:
\begin{equation}
F(x)=f_0+\frac{f_1}{x}+\frac{f_2}{x^2}+{\cal O}(x^{-3})
\end{equation}
and
\begin{equation}
G(x)= (n-3)^2 x^2+g_1 x+g_0.
\end{equation}
where $f_0$, $f_1$, $f_2$, $g_0$, and $g_1$ are constants.
Two different asymptotic behaviors  appear depending whether or not $f_0$ vanishes.

\subsubsection{Asymptotically flat solutions}
First we consider the case of $f_0 \neq 0 $ and then define a new time coordinate $T$ and the areal coordinate $R$ by
\begin{equation} \label{Rareal}
R=\left(F^2 G \right)^{1/\{2(n-3)\}}, \qquad T=\frac{t}{f_0},
\end{equation}
so that  the line element \eqref{gauge-higher} becomes
\begin{equation} \label{TR}
\D s^2=g_{TT} \D T^2+\frac{\D R^2}{g^{RR}}+R^2 \D\Omega^2_{n-2}.
\end{equation}
The relation between $x$ and $R$ for $R\to \infty$ is
\begin{align}
x(R)=&\frac{R^{n-3} }{(n-3)f_0 }-\frac{ f_0 g_1 +2 (n-3)^2f_1 }{2 (n-3)^2f_0 } \nonumber \\&+\frac{f_0 \left\{g_1^2-4 (n-3)^2g_0 \right\}-4  (n-3)^2 f_1 g_1-8  (n-3)^4f_2 }{8 (n-3)^3 R^{n-3} }+{\cal O}\left(\frac{1}{R^{2(n-3)}}\right),
\end{align}
which shows
\begin{align}
-g_{T T}=&1-\frac{2  (n-3)f_1}{R^{n-3} }+\frac{(n-3)^2(f_1^2 -2 f_0 f_2) -f_0 f_1 g_1}{R^{2(n-3)} }+{\cal O}\left(\frac{1}{R^{3(n-3)}}\right), \label{gTT}\\
g^{R R}=&1-\frac{2  (n-3)f_1}{R^{n-3} } \nonumber \\
&+\frac{f_0^2 g_1^2+4 (n-3)^4\left(f_1^2-4 f_0 f_2\right)-4 (n-3)^2f_0(f_0 g_0+2 f_1 g_1)}{4 (n-3)^2R^{2(n-3)}}+{\cal O}\left(\frac{1}{R^{3(n-3)}}\right)\label{gRR}
\end{align}
for $R\to \infty$.
Thus, it is concluded that the solutions I--IV with $k=1$ and the solution V${}_1$ are asymptotically flat for $x\to \infty$ provided $f_0 \neq 0$.

Additionally, the asymptotic form of the electric and scalar field in this coordinate system are
\begin{align} 
F_{R T}=& \frac{q}{R^{n-2} }-\frac{q f_0\left[f_0 \left\{g_1^2-4 (n-3)^2 g_0\right\}-4 (n-3)^2 f_1 g_1-8 (n-3)^4  f_2\right]}{8 (n-3)^2 R^{3n-8} } \nonumber \\
&\qquad \qquad \qquad \qquad \qquad \qquad \qquad \qquad \qquad \qquad \qquad \qquad +{\cal O}\left(\frac{1}{R^{4n-11}}\right), \label{FRT} \\
\phi(R)=&\phi_0-\frac{\phi_1 f_0}{(n-3)R^{n-3} }-\frac{ \phi_1f_0 f_1}{ R^{2(n-3)}}+{\cal O}\left(\frac{1}{R^{3(n-3)}}\right), \label{phiR}
\end{align}
where the relations
\begin{align}
\kappa_{n} \phi_1^2=&\frac{(n-2) \left[f_0 \left\{g_1^2-4 (n-3)^2g_0 \right\}-4 (n-3)^2 f_1 g_1-8 (n-3)^4f_2 \right]}{4 (n-3) f_0},\\
\kappa_{n} q^2=&(n-2)(n-3)  \left\{(f_1^2-2 f_0 f_2 )(n-3)^2-f_0 f_1 g_1\right\}
\end{align}
are obtained from the asymptotic field equations.

\subsubsection{Asymptotically Bertotti-Robinson solutions}
In the case of $f_0=0$, on the other hand, the spacetime is not asymptotically flat for $x\to \infty$.
This is because the areal coordinate (\ref{Rareal}) converges to a constant as
\begin{align}
\lim_{x\to \infty}R=\lim_{x\to \infty} \left(F^2 G \right)^{1/\{2(n-3)\}}=\left\{(n-3)^2f_1^2\right\}^{1/\{2(n-3)\}}=:R_0.\label{limR-BR}
\end{align}
Moreover, for large $x$ with $f_0=0$, the leading terms of $g_{tt}$ and $g^{xx}$ behave as  $x^2$ and both the scalar and the electric field converge to a constant. 
The asymptotic behavior for $x\to \infty$ are explicitly given as
\begin{align}
&\lim_{x\to \infty}\D s^2 \simeq -\frac{x^2}{f_1^2}\D { t}^2+\frac{f_1^2}{R_0^{2(n-4)}}\frac{\D x^2}{x^2}+R_0^2\D\Omega^2_{n-2},\label{f0asymp}\\
&\lim_{x\to \infty}F_{xt}=\frac{q}{R_0^{2(n-3)}}, \qquad \lim_{x\to \infty}\phi=\phi_0.
\end{align}
Thus, the solutions I--IV with $k=1$ and the solution V${}_1$ approach the higher-dimensional Bertotti-Robinson solution for  $x\to \infty$ in the case $f_0=0$.

Here we show that $x\to \infty$ corresponds to null infinity.
Let us consider an affinely-parametrized radial null geodesic $x^\mu(\lambda)=(t(\lambda),x(\lambda),0,\cdots,0)$, where $\lambda$ is an affine parameter.
Such a geodesic satisfies
\begin{align}
0=-F^{-2}{\dot t}^2+F^{2/(n-3)}G^{-(n-4)/(n-3)}{\dot x}^2,\label{geodesics1}
\end{align}
where a dot denotes differentiation with respect to $\lambda$.
Also, along such a geodesic, $C_{(t)}:=k_\mu \xi^\mu_{(t)}=-F^{-2}{\dot t}$ is a conserved quantity associated with the Killing vector ${\xi}^\mu_{(t)}=(1,0,\cdots,0)$, where $k^\mu=({\dot t},{\dot r},0,\cdots,0)$ is the tangent vector of the geodesic.
Thus from Eq.~(\ref{geodesics1}), we obtain
\begin{align}
{\dot x}^2=C_{(t)}^2F(x)^{2(n-4)/(n-3)}G(x)^{(n-4)/(n-3)}.\label{geodesics2}
\end{align}
Because the right-hand side of the above equation reduces to a constant for $x\to\infty$ as
\begin{align}
\lim_{x\to\infty}C_{(t)}^2F(x)^{2(n-4)/(n-3)}G(x)^{(n-4)/(n-3)}=C_{(t)}^2R_0^{2(n-4)}
\end{align}
by Eq.~(\ref{limR-BR}), we obtain
\begin{align}
x(\lambda)\simeq \pm C_{(t)}R_0^{n-4}\lambda
\end{align}
for $x\to \infty$.
Since the affine parameter $\lambda$ blows up for $x\to\infty$, it corresponds to null infinity.

Different from the asymptotically flat case, null infinity $x\to\infty$ is timelike in the asymptotically Bertotti-Robinson solutions.
This is because the two-dimensional Lorentzian portion of the Bertotti-Robinson spacetime (\ref{f0asymp}) is AdS${}_2$.

\subsection{Conserved charges }
Here we use the Regge-Teitelboim method~\cite{Regge:1974zd} for computing the conserved charges--mass and electric charge--of the asymptotically flat and spherically symmetric solutions. In this Hamiltonian based approach, the mass $M$ is given by a surface integral and it corresponds to the conserved charge associated with the time translation symmetry of a static configuration. A scalar field can add a non-vanishing term to the surface integral which contributes to the mass~\cite{Henneaux:2002wm,Gegenberg:2003jr,Henneaux:2004zi,Henneaux:2006hk}.
As expected, the mass depends on the boundary conditions imposed on the fields. In fact, for a massless scalar field \eqref{phiR}, the contribution to the surface integral vanishes if the boundary condition $\delta \phi_0=0$ is imposed. (See Eq.~(24) in \cite{Saenz:2012ga}.) Thus, if $\phi_0$ is a fixed constant, the mass is just given by the surface integral corresponding to pure gravity (Eq.~(23) in~\cite{Saenz:2012ga}). 
Hereafter, we assume that boundary condition. 

For asymptotically flat spacetimes in pure gravity, or in those cases where there are no additional surface integrals contributing the the mass,  it is possible to determine the mass $M$ from the coefficient $m$ of the sub-leading term $1/R^{n-3}$ of $g^{RR}$ in the spherically symmetric metric \eqref{TR},  which is given by
\begin{equation}
M=\frac{(n-2) V_{n-2}}{2 \kappa_{n}} m,
\end{equation}
where $V_{n-2}$ is the volume of $S^{n-2}$.
In our case, the coefficient $m$ in the asymptotic form of $g^{RR}$, given by \eqref{gRR}, is
\begin{equation}
m=2 (n-3) f_1.
\end{equation}

In what follows we show that he electric charge $Q$ is related to integration constant  $q$ as
\begin{equation}
Q= q  V_{n-2}.
\end{equation}
In the Hamiltonian formalism, the line element is written in the following form
\begin{equation}
\D s^2=-(N^{\perp})^2\D t^{2}+h _{ij}\left( N^{i}\D t+\D x^{i}\right) \left(N^{j}\D t+\D x^{j}\right),
\end{equation}
where the metric coefficients in terms of  $h _{ij}$, $N^{\perp}$, $N^{i}$ are
\begin{eqnarray}
g_{tt} &=&-(N^{\perp})^2+g_{ij}N^{i}N^{j},  \notag \\
g_{ti} &=&h_{ij}N^{j}=N_{i},    \label{CompCov} \\
g_{ij} &=&h_{ij},  \notag
\end{eqnarray}
and the inverse metric is given by 
\begin{eqnarray}
g^{tt} &=&-(N^{\perp})^{-2},  \notag \\
g^{ti} &=&N^{i}(N^{\perp})^{-2},   \label{CompCont} \\
g^{ij} &=&h^{ij}-N^{i}N^{j}(N^{\perp})^{-2}.  \notag
\end{eqnarray}
The determinant of the metric is  $\sqrt{-g}=N^{\perp}\sqrt{h}$, where $h:= \det(h_{ij})$.

The Hamiltonian for the Maxwell Lagrangian,
\begin{equation}
-\frac{1}{4} \int \D^nx \sqrt{-g}  F_{ \mu \nu}  F^{ \mu \nu},
\end{equation}
is a function of the canonical variables, which are the spatial components of the gauge field $A_i$ and their corresponding momenta $P^i$ given by
\begin{equation}
P^i=N^{\perp} \sqrt{h} F^{i t}=\sqrt{h}(N^{\perp})^{-1} h^{i j}\left(  F_{t j}+N^kF_{j k} \right).
\end{equation}
The electric charge $Q$ is the conserved charge associated with the gauge symmetry generated by the Gauss constraint $\partial_i P^i=0$ and given by
\begin{equation}
Q=-\oint_{S^{n-2}} \D S_i P^i,\label{Q-charge-def}
\end{equation}
where the integral is performed on $S^{n-2}$ at spacelike infinity.
A minus sign is included in the above definition because the electric field $F^{t i}$ and the canonical momentum of the gauge field $P^i$ have different signs.

In our case, $N^i=0$ holds so that Eq.~(\ref{Q-charge-def}) gives
\begin{equation}
Q=\lim_{x\to \infty}\oint_{S^{n-2}}  \D S_x\sqrt{h}(N^{\perp})^{-1} h^{xx}  F_{xt}.
\end{equation}
Finally, using $\D S_x= \D z^1 \cdots \D z^{n-2}$, $N^{\perp}=F^{-1}$, $F_{xt}= q F^{-2} G^{-1}$, and $\sqrt{h}=h_{xx} F G \sqrt{\gamma}$, where $\gamma = \det(\gamma_{ab})$, we obtain
\begin{align}
Q= q \oint \D z^1 \cdots \D z^{n-2} \sqrt{\gamma} = q V_{n-2}.
\end{align}

In the terms of the conserved charges and the amplitude of the scalar field $\phi_1$, the asymptotic form of the solutions I--IV with $k=1$ and the solution V${}_1$ is given by
\begin{align}
-g_{T T}=&1-\frac{m}{R^{n-3} }+\frac{\kappa_{n}  q^2}{(n-2) (n-3) R^{2(n-3)} }+{\cal O}\left(\frac{1}{R^{3(n-3)}}\right),\\
g^{R R}=&1-\frac{m}{R^{n-3} }+\frac{\kappa_{n}  q^2+\kappa_{n}  \phi_1^2 f_0^2}{(n-2) (n-3) R^{2(n-3)} }+{\cal O}\left(\frac{1}{R^{3(n-3)}}\right),\\
\phi(R)=&\phi_0-\frac{\phi_1 f_0}{(n-3)R^{n-3} }-\frac{m  \phi_1f_0}{2 (n-3) R^{2(n-3)}}+{\cal O}\left(\frac{1}{R^{3(n-3)}}\right),\\
F_{R T}=& \frac{q}{R^{n-2} }+{\cal O}\left(\frac{1}{R^{3n-8}}\right).
\end{align}

\subsection{Causal structure}
We have shown that asymptotic flatness imposes a constraint on the parameters in the solutions I--IV and V${}_1$.
Now let us clarify the global structure of those asymptotically flat spacetimes.
By the no-hair theorem, none of the solutions with a non-trivial scalar field admit a Killing horizon.
As a result, the domain of $x$ for an asymptotically flat spacetime is given by $x_0\le x<\infty$, where $x=x_0$ is the location of a curvature singularity.

In our coordinate system (\ref{gauge-higher}), the structure of the singularity in the Penrose diagram is clarified by the two-dimensional portion of the spacetime $M^2$, of which line-element $\D s_2^2$ is 
\begin{align} \label{ds2two}
\D s_2^2=&F(x)^{-2}\left\{-\D t^2+F(x)^{2(n-2)/(n-3)}G(x)^{-(n-4)/(n-3)}\D x^2\right\}.
\end{align}
The line-element $\D {\bar s}_2^2$ of the conformally completed spacetime ${\bar M}^2$ given by $\D {\bar s}_2^2:=F(x)^{2}\D s_2^2$ is cast into the following form:
\begin{align}
\D {\bar s}_2^2=-\D t^2+\D {\bar r}^2
\end{align}
by introducing a new radial coordinate ${\bar r}$ defined by 
\begin{align}
{\bar r}:=\int^x F(x)^{(n-2)/(n-3)}G(x)^{-(n-4)/\{2(n-3)\}}\D x.
\end{align}
A hypersurface with constant $x$ is timelike if it corresponds to a finite value of ${\bar r}$, while it is null if it corresponds to ${\bar r}\to \pm\infty$.

\begin{figure}[htbp]
\begin{center}
\includegraphics[width=0.2\linewidth]{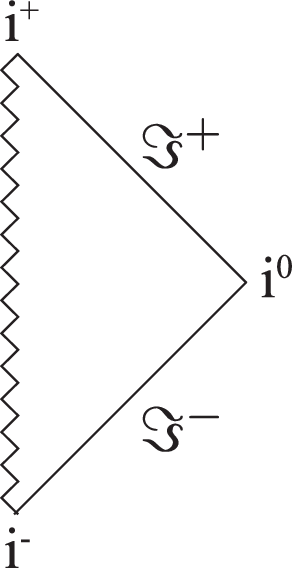}
\caption{\label{fig1} The Penrose diagram for all the asymptotically flat spacetimes given by the solutions I--IV and V${}_1$ with nontrivial configurations of a scalar field and a Maxwell field.
A zigzag line represents a timelike curvature singularity.
$\Im^{+(-)}$ corresponds to the future (past) null infinity.
$i^{+(-)}$ corresponds to the future (past) timelike infinity, while $i^0$ is the spacelike infinity. }
\end{center}
\end{figure}

In this subsection, we will show that all the curvature singularities in the asymptotically flat spacetimes,  represented by the solutions I--IV and V${}_1$ with nontrivial configurations of a scalar field and a Maxwell field, are timelike.
As a result, Fig.~\ref{fig1} is the Penrose diagram for all of them.

\subsubsection{Type-I solution}
The metric functions $G(x)$ and $F(x)$ for the type-I solution are given by Eqs.~(\ref{sol-G1}) and (\ref{sol-F1}) with $k=1$ and $\varepsilon=1$, respectively:
\begin{align}
G(x)=&(n-3)^2(x-a)(x-b),\\
F(x)=& A\biggl(\frac{x-a}{x-b}\biggl)^{\alpha/2}+B\biggl(\frac{x-a}{x-b}\biggl)^{-\alpha/2}.
\end{align}
We assume reality of the scalar field and then $0<\alpha<1$ without loss of generality.
We also assume $AB(a-b) \ne 0$ for a nontrivial Maxwell field and $A\ne -B$ for asymptotic flatness.

Both $x=a$ and $x=b$ correspond to curvature singularities where a scalar field diverges.
In addition, $x=x_{\rm s}$ defined by $F(x_{\rm s})=0$ is also a curvature singularity but with a finite scalar field, where $x_{\rm s}$ is given by 
\begin{align}
x_{\rm s}=\frac{a-b(-B/A)^{1/\alpha}}{1-(-B/A)^{1/\alpha}}.
\end{align}
The above expression shows $x_{\rm s}>a>b$ or $b>a>x_{\rm s}$ for $0<(-B/A)^{1/\alpha}<1$ and $x_{\rm s}>b>a$ or $a>b>x_{\rm s}$ for $(-B/A)^{1/\alpha}>1$.
The largest value among $x_{\rm s}$, $a$, and $b$ corresponds to the singularity which appears in the asymptotically flat spacetime.

In the limit $x\to x_{\rm s}$, we have
\begin{align}
\lim_{x\to x_{\rm s}}F(x)^{(n-2)/(n-3)}G(x)^{-(n-4)/\{2(n-3)\}}=0
\end{align}
and hence the singularity $x=x_{\rm s}$ is timelike.
In the limit $x\to a$, we have
\begin{align}
\lim_{x\to a}F(x)^{(n-2)/(n-3)}G(x)^{-(n-4)/\{2(n-3)\}}\propto (x-a)^p,
\end{align}
where 
\begin{align}
p:=-\frac{\alpha(n-2)+(n-4)}{2(n-3)}.
\end{align}
The singularity $x=a$ is timelike and null for $p>-1$ and $p\le -1$, respectively.
Thus, for $0< \alpha<1$, the singularity $x=a$ is timelike.
This argument is also valid for $x=b$ and hence it is timelike.

\subsubsection{Type-II solution}
We assume $(a-b){\bar \alpha}\ne 0$ in the type-II solution for a nontrivial Maxwell field, of which metric functions $G(x)$ and $F(x)$ are given by Eqs.~(\ref{sol-G1}) and (\ref{type-I-g1000}) with $k=1$ and $\varepsilon=1$, respectively:
\begin{align}
G(x)=&(n-3)^2(x-a)(x-b),\\
F(x)=&F_1\cos\biggl\{\frac{{\bar \alpha}}{2}\ln\biggl(\frac{x-a}{x-b}\biggl)\biggl\}+F_2\sin\biggl\{\frac{{\bar \alpha}}{2}\ln\biggl(\frac{x-a}{x-b}\biggl)\biggl\}.
\end{align}
We also assume $F_1\ne 0$ for asymptotic flatness.

While both $x=a$ and $x=b$ correspond to a curvature singularity where the scalar field diverges, $x=x_{\rm s}$ defined by $F(x_{\rm s})=0$ is a curvature singularity with a finite scalar field.
In the type-II solution, $x_{\rm s}$ is multiple characterized by an integer $N$;
\begin{equation} 
x_{\rm s}(N)= \left\{ \begin{array}{ll}
\displaystyle \frac{a-b\exp\{2(2N\pi-\theta_0)/{\bar \alpha}\}}{1-\exp\{2(2N\pi-\theta_0)/{\bar \alpha}\}}  & \mbox{for $F_2\ne 0$},\\[4mm]
\displaystyle \frac{a-b\exp\{(2N+1)\pi/{\bar \alpha}\}}{1-\exp\{(2N+1)\pi/{\bar \alpha}\}} &        \mbox{for $F_2=0$}
        ,\end{array} \right.
\end{equation}
where $\theta_0:=\arctan(F_1/F_2)$ and $N$ is an integer.
The largest value among $x_{\rm s}$, $a$, and $b$ corresponds to the singularity which appears in the asymptotically flat spacetime.

We can show that there is always a value of $N$ such that $x_{\rm s}(N)>\max\{a,b\}$ independent of the parameters, namely there is at least one curvature singularity $x=x_{\rm s}$ which is located outside the singularities at $x=a$ and $x=b$.
This is shown by the fact that $x_{\rm s}$ has the following form:
\begin{equation} 
x_{\rm s}=\frac{a-bw}{1-w},
\end{equation}
where $w$ takes discrete values in the domain $w>0$ and $w\to \infty$ and $w\to 0$ are realized for $N\to \infty$ or $N\to -\infty$ depending on the parameters.
From the above expression, we obtain
\begin{equation} 
x_{\rm s}-a=\frac{(a-b)w}{1-w},\qquad x_{\rm s}-b=\frac{a-b}{1-w}.
\end{equation}
While we have $x_{\rm s}-a\to b-a$ and $x_{\rm s}-b\simeq (b-a)/w$ for large $w$, we have $x_{\rm s}-a\simeq (a-b)w$ and $x_{\rm s}-b= a-b$ for small $w$.
Therefore, independent on the sign of $a-b$, there always exists a value of $N$ such that $x_{\rm s}(N)>a$ and $x_{\rm s}(N)>b$ hold.

In the limit $x\to x_{\rm s}$, we have
\begin{align}
\lim_{x\to x_{\rm s}}F(x)^{(n-2)/(n-3)}G(x)^{-(n-4)/\{2(n-3)\}}=0
\end{align}
and hence the singularity $x=x_{\rm s}$ is timelike.
On the other hand, the value of the metric function $F(x$) is indefinite for $x\to a$ or $x\to b$ because of its oscillatory nature.
A more careful study is necessary to clarify the signature of the curvature singularities at $x\to a$ and $x\to b$.
We leave this problem for future investigations.

\subsubsection{Type-III solution}
The metric functions $G(x)$ and $F(x)$ of the type-III solution are given by Eqs.~(\ref{sol-G1}) and (\ref{sol-F2}) with $k=1$ and $\varepsilon=1$, respectively:
\begin{align}
G(x)=&(n-3)^2(x-a)(x-b),\\
F(x)=& A\ln\biggl(\frac{x-a}{x-b}\biggl)+B.
\end{align}
We assume $A(a-b)\ne 0$ to give a nontrivial Maxwell field and $B\ne 0$ for asymptotic flatness.

Both $x=a$ and $x=b$ correspond to curvature singularities where a scalar field diverges.
Also, $x=x_{\rm s}$ defined by $F(x_{\rm s})=0$ is a curvature singularity but with a finite scalar field, where $x_{\rm s}$ is given by 
\begin{align}
x_{\rm s}=\frac{a-be^{-B/A}}{1-e^{-B/A}}.
\end{align}
This expression shows $x_{\rm s}>a>b$ or $b>a>x_{\rm s}$ for $0<e^{-B/A}<1$ and $x_{\rm s}>b>a$ or $a>b>x_{\rm s}$ for $e^{-B/A}>1$.
The largest value among $x_{\rm s}$, $a$, and $b$ corresponds to the singularity which appears in the asymptotically flat spacetime.

In the limit $x\to x_{\rm s}$, we have
\begin{align}
\lim_{x\to x_{\rm s}}F(x)^{(n-2)/(n-3)}G(x)^{-(n-4)/\{2(n-3)\}}=0,
\end{align}
and hence the singularity $x=x_{\rm s}$ is timelike.
In the limit $x\to a$, we have
\begin{align}
\lim_{x\to a}F(x)^{(n-2)/(n-3)}G(x)^{-(n-4)/\{2(n-3)\}}\propto (x-a)^{-(n-4)/\{2(n-3)\}}\left\{\ln(x-a)\right\}^{(n-2)/(n-3)}.
\end{align}
The right-hand side blows up slower than $(x-a)^{-\epsilon-[(n-4)/\{2(n-3)\}]}$, where $\epsilon$ is a positive number satisfying $0<\epsilon<(n-2)/\{2(n-3)\}$.
Since the divergence $(x-a)^{-\epsilon-[(n-4)/\{2(n-3)\}]}$ corresponds to finite ${\bar r}$ and therefore the singularity $x=a$ is timelike.
This argument is also valid for $x=b$ and hence it is timelike.

\subsubsection{Type-IV solution}
We assume $\phi_1\ne 0$ in the type-IV solution for a nontrivial Maxwell field, of which metric functions $G(x)$ and $F(x)$ are given by Eqs.~(\ref{sol-G1}) and (\ref{sol-F3}) with $k=1$, respectively:
\begin{align}
G(x)=&(n-3)^2(x-a)^2,\\
F(x)=&A\sin\biggl(\sqrt{\frac{\kappa_{n}}{(n-2)(n-3)^3}}\frac{\phi_1}{x-a}\biggl)+B\cos\biggl(\sqrt{\frac{\kappa_{n}}{(n-2)(n-3)^3}}\frac{\phi_1}{x-a}\biggl).
\end{align}
We also assume $B\ne 0$ for asymptotic flatness.

While $x=a$ corresponds to a curvature singularity where a scalar field diverges, $x=x_{\rm s}$ defined by $F(x_{\rm s})=0$ is a curvature singularity with a finite scalar field.
In the type-IV solution, $x_{\rm s}$ is multiple and given by
\begin{equation} 
x_{\rm s}= \left\{ \begin{array}{ll}
\displaystyle a+\sqrt{\frac{\kappa_{n}}{(n-2)(n-3)^3}}\frac{\phi_1}{\arctan(-B/A)}  & \mbox{for $A\ne 0$},\\[4mm]
\displaystyle a+\sqrt{\frac{\kappa_{n}}{(n-2)(n-3)^3}}\frac{2\phi_1}{(2N+1)\pi} &        \mbox{for $A=0$}
        ,\end{array} \right.
\end{equation}
where $N$ is an integer.
Because there is an infinite number of values of $x_{\rm s}$ greater than $a$, the largest value of $x_{\rm s}$ corresponds to the singularity which appears in the asymptotically flat spacetime.
In the limit of $x=x_{\rm s}$, we have
\begin{align}
\lim_{x\to x_{\rm s}}F(x)^{(n-2)/(n-3)}G(x)^{-(n-4)/\{2(n-3)\}}=0
\end{align}
and hence the singularity $x=x_{\rm s}$ is timelike.

\subsubsection{Type-V${}_1$ solution}
In the type-V${}_1$ solution, where $G_0>0$ is assumed, the metric functions $G(x)$ and $F(x)$ are given by Eqs.~(\ref{sol-G2}) and (\ref{sol-F4}) with $k=1$, respectively:
\begin{align}
G(x)=&(n-3)^2x^2+G_0,\\
F(x)=&A\sin\biggl\{\sqrt{\frac{\kappa_{n}\phi_1^2+(n-2)(n-3)G_0}{(n-2)(n-3)G_0}}\arctan\biggl(\frac{(n-3)x}{\sqrt{G_0}}\biggl)\biggl\} \nonumber \\
+&B\cos\biggl\{\sqrt{\frac{\kappa_{n}\phi_1^2+(n-2)(n-3)G_0}{(n-2)(n-3)G_0}}\arctan\biggl(\frac{(n-3)x}{\sqrt{G_0}}\biggl)\biggl\}.
\end{align}
We also assume $\phi_1\ne 0$ for nontrivial scalar field and 
\begin{equation} 
\frac{\pi}{2}\ne  \left\{ \begin{array}{ll}
\displaystyle \sqrt{\frac{(n-2)(n-3)G_0}{\kappa_{n}\phi_1^2+(n-2)(n-3)G_0}}\arctan\biggl(-\frac{B}{A}\biggl) & \mbox{for $A\ne 0$},\\[4mm]
\displaystyle \frac{(2N+1)\pi}{2}\sqrt{\frac{(n-2)(n-3)G_0}{\kappa_{n}\phi_1^2+(n-2)(n-3)G_0}} &        \mbox{for $A=0$},\\[4mm]
\displaystyle N\pi\sqrt{\frac{(n-2)(n-3)G_0}{\kappa_{n}\phi_1^2+(n-2)(n-3)G_0}} &        \mbox{for $B=0$}
        ,\end{array} \right.
\end{equation}
for asymptotic flatness, where $N$ is an integer.

The location of the singularity $x=x_{\rm s}$ satisfying $F(x_{\rm s})=0$ is given by 
\begin{equation} 
x_{\rm s}= \left\{ \begin{array}{ll}
\displaystyle \frac{\sqrt{G_0}}{n-3}\tan\biggl\{\sqrt{\frac{(n-2)(n-3)G_0}{\kappa_{n}\phi_1^2+(n-2)(n-3)G_0}}\arctan\biggl(-\frac{B}{A}\biggl)\biggl\} & \mbox{for $A\ne 0$},\\[4mm]
\displaystyle \frac{\sqrt{G_0}}{n-3}\tan\biggl\{\frac{(2N+1)\pi}{2}\sqrt{\frac{(n-2)(n-3)G_0}{\kappa_{n}\phi_1^2+(n-2)(n-3)G_0}}\biggl\} &        \mbox{for $A=0$},\\[4mm]
\displaystyle \frac{\sqrt{G_0}}{n-3}\tan\biggl\{N\pi\sqrt{\frac{(n-2)(n-3)G_0}{\kappa_{n}\phi_1^2+(n-2)(n-3)G_0}}\biggl\} &        \mbox{for $B=0$}
        ,\end{array} \right.
\end{equation}
where $N$ is an integer.
Similar to the type-IV solution, there is an infinite number of values of $x_{\rm s}$, all of which correspond to finite $\phi$.
In the limit of $x=x_{\rm s}$, we have
\begin{align}
\lim_{x\to x_{\rm s}}F(x)^{(n-2)/(n-3)}G(x)^{-(n-4)/\{2(n-3)\}}=0
\end{align}
and hence all the singularities $x=x_{\rm s}$ are timelike.


\section{Concluding remarks}
In the present paper, we have presented a complete set of static solutions in the Einstein-Maxwell system with a non-constant massless scalar field in arbitrary $n(\ge 3)$ dimensions.
We have considered warped product spacetimes $M^2 \times K^{n-2}$, where $K^{n-2}$ is a $(n-2)$-dimensional Einstein space and assumed that the scalar field depends only on the radial coordinate and the electromagnetic field is purely electric. 

While the solution is unique in any dimensions in the absence of a Maxwell field, there are multiple distinct solutions with a nontrivial Maxwell field.
The general solution consists of nine solutions for $n\ge 4$ and three solutions for $n=3$, which are all written by elementary functions and summarized in Table~\ref{Table:solutions}. 
None of them are endowed of a Killing horizon in accordance with the no-hair theorem.
The solutions in four and higher dimensions are also obtained in a different but useful coordinate system, which are presented in Appendix~\ref{App:gauge}.

We have clarified limiting cases where the Maxwell field or scalar field is trivial and identified solutions which represent asymptotically flat spacetimes depending on the integration constants.
In three dimensions, there is a unique asymptotically locally flat solution only in the case without a Maxwell field.
In four and higher dimensions, while the asymptotically flat solution is unique with a vanishing Maxwell field or constant scalar field, there are five different asymptotically flat solutions in the case with nontrivial configurations of a Maxwell field and a scalar field. Moreover, the solutions for $n\ge 4$, which include the arbitrary dimensional generalization of the Penney solution, can be asymptotically Bertotti-Robinson spacetime depending on the integration constants.

Along the text we have considered a real scalar field.  
However, one can consider also a phantom scalar field. 
This case follows from our solutions by including the condition $\phi_1^2 <0$. In Table 1 we have pointed out the solutions allowing a phantom scalar field, which remain real after direct  analytic extensions. Nevertheless, in 
 such a case with a phantom scalar field, a complete classification requires an additional solution in Section~\ref{sec:noroot}.
The general solution in the case of $\kappa_{n}\phi_1^2= -(n-2)(n-3)kG_0(<0)$ is given by 
\begin{align}
F(x)=&A\arctan\biggl(\frac{(n-3)kx}{\sqrt{kG_0}}\biggl)+B,\label{sol-F5}
\end{align}
where
\begin{align}
A^2=\frac{\kappa_{n} q^2}{(n-2)(n-3)kG_0 },
\end{align}
and $B$ is an integration constant. This corresponds to an arbitrary dimensional generalization of the Ellis wormhole~\cite{Ellis1973} with a Maxwell field.
This solution with $n=4$ and $k=1$ was given in~\cite{ggs2009}.

One of the possible generalization of the present work is to add a cosmological constant. Even without the Maxwell field, a complete classification of the static solutions has not been performed yet. Only the general solution for the case of Ricci flat base manifolds ($k=0$), and in presence of a negative cosmological constant, is known in any spacetime dimension \cite{Saenz:2012ga}.
Another possible generalization is to consider the dilatonic coupling of the scalar field to the Maxwell field. Results in this direction, following a  similar approach to~\cite{pk1996}, can be found in \cite{kp1997}. 
We will address these problems elsewhere.

\subsection*{Acknowledgements}
HM thanks Marco Astorino for valuable comments. 
HM also thanks the Theoretical Physics group in CECs and Universidad Adolfo Ib\'a\~{n}ez for hospitality and support, where this work was completed.
This work has been partially funded by the Fondecyt
grants 1121031, 1130658, 1161311 and 1180368. The Centro de Estudios Cient\'{\i}ficos (CECs) is funded by the Chilean Government through the Centers of Excellence Base Financing Program of Conicyt.

\appendix

\section{Another useful gauge in four and higher dimensions}
\label{App:gauge}
\subsection{Basic equations}
In four and higher dimensions we consider a new radial coordinate $r$, such as $\D x= G \D r$. In this gauge the metric (\ref{gauge-higher}) reads now 
\begin{align}
\D s^2=&-F(r)^{-2}\D t^2+F(r)^{2/(n-3)}G(r)^{1/(n-3)}\biggl(G(r)\D r^2+\gamma_{ab}(z)\D z^a\D z^b\biggl),\label{gauge-higher-other}
\end{align}
so that the field equations (\ref{em-kg}) yield
\begin{align}
F_{rt}=\frac{q}{F^{2}},\qquad \frac{\D \phi}{\D r}=c,
\end{align}
where $c$ and $q$ are constants.
Now we write the Einstein equations as ${\cal E}^\mu_{~~\nu}=0$, where ${\cal E}^\mu_{~~\nu}$ is defined by Eq.~(\ref{EFE-different}).

Using the gauge (\ref{gauge-higher-other}), and defining (for $n\ge 4$)
\begin{equation} \label{rede}
F=e^{-b} \quad  \mbox{and} \quad  G= h^{-2},
\end{equation}
we obtain
\begin{align}
\mathcal{E}_{t} ^{t}=0 &\Rightarrow b''-\frac{n-3}{n-2}\kappa_{n} q^2 e^{2 b} = 0,  
\label{tt0}\\
\mathcal{E}_{r} ^{r}=0 &\Rightarrow h''-\frac{h}{n-2}\left\{(n-3)\biggl( \kappa_{n} c^2
-\frac{n-3}{n-2}\kappa_{n} q^2 e^{2 b} \biggl) +(n-2)b'^2-b''\right\} = 0, \label{rr0}\\
\mathcal{E}_{j} ^{i}=0 &\Rightarrow \left\{(n-3)^2 k-h'^2+\biggl(b''
-\frac{n-3}{n-2}\kappa_{n} q^2 e^{2 b}\biggl)h^2+hh''\right\}\delta_{j} ^{i}  = 0, \label{ij0}
\end{align}
where a prime denotes the derivative with respect to $r$.
Replacing (\ref{tt0}) in (\ref{rr0}) and (\ref{ij0}), and defining the constants
\begin{equation}
 q_n^2:=2\frac{n-3}{n-2}q^2,  \quad
c_n^2:=2\frac{n-3}{n-2} c^2, \quad 
 \gamma_n:=(n-3)^2 k,
\end{equation}
a simple system of differential equations is obtained:
\begin{align}
 b''-\frac12\kappa_n q_n^2 e^{2 b} &= 0,  
\label{tt}\\
 h''-\left(\frac12\kappa_n c_n^2 
+b'^2-b''\right)h &= 0, \label{rr}\\
 hh''-h'^2+\gamma_n  &= 0. \label{ij}
\end{align}
Remarkably, this system in arbitrary dimensions $n\ge 4$ exactly takes the same form as in four dimensions.

A first integral of Eq. (\ref{tt}) can be obtained by setting $b''= b'\D b'/\D b$. Thus, we have
 \begin{equation} \label{tt2}
 b'^2= \frac{1}{2}\kappa_{n} q_n^2 e^{2 b} +b_1,
 \end{equation}
where $b_1$ is an integration constant.  From (\ref{tt}) and (\ref{tt2}) we note that $b'^2-b''=b_1$, which reduces (\ref{rr}) to
$ h''-a_1 h=0 $ with $a_1=\kappa_{n} c_n^2/2+b_1$.

In summary, the system to be solved takes a very simple form: 
\begin{align}
b'^2- \frac{1}{2}\kappa_{n} q_n^2 e^{2 b} -b_1&= 0,  
\label{tt3}\\
  h''-a_1 h &= 0, \label{rr2}\\
a_1 h^2-h'^2+\gamma_n  &= 0. \label{ij2}
\end{align}

\subsection{Solutions}

Equation (\ref{tt3}) is easily solved by direct integration yielding
\begin{equation} \label{FApend}
F^{-2}= e^{2 b}=\left\{ \begin{array}{ll}
\displaystyle \frac{2}{\kappa_{n} q_n^2} (r-r_0)^{-2} & \mbox{if $b_1= 0$},\\[4mm]
\displaystyle \frac{2 b_1}{\kappa_{n} q_n^2} \left(\sinh{\sqrt{b_1}(r-r_0)}\right)^{-2} & 
       \mbox{if $b_1 > 0$}, \\[4mm]
\displaystyle -\frac{2 b_1}{\kappa_{n} q_n^2}\left(\sin{\sqrt{-b_1}(r-r_0)}\right)^{-2}
       & \mbox{if $b_1 < 0$}, \\[4mm]
\displaystyle \exp \biggl(2\sqrt{b_1}(r-r_0)\biggl)
       & \mbox{if $q_n =0$}
        .\end{array} \right.
\end{equation}

The solution of (\ref{rr2}) depends on the sign of  $a_1$. Thus,  the following cases appear:
\subsubsection{$a_1 > 0$}
This occurs if $b_1>-\kappa_{n} c_n^2/2$. In this case (\ref{rr2}) gives
\begin{equation}
h=c_1e^{\sqrt{a_1}r}+c_2e^{-\sqrt{a_1}r},
\end{equation}
where the integration constants $c_1, c_2$ are constrained by (\ref{ij2}) to hold
\begin{equation}
4 a_1 c_1 c_2 + \gamma_n=0.
\end{equation}
Note that if $k=0$, and hence  $\gamma_n=0$, one of the constants  $c_1, c_2$ must be 0.
\subsubsection{$a_1 = 0$}
This is the case when $b_1=-\kappa_{n} c_n^2/2$. Here, the general solution of (\ref{rr2}) is
\begin{equation}
h=c_1 r+c_2.
\end{equation}
Now, from (\ref{ij2}) we note that the integration constant $c_1$ must satisfy $c_1^2=\gamma_n$. Then, this case is not possible for $k=-1$. 
\subsubsection{$a_1 < 0$}
The constant $a_1$ is negative if $b_1<-\kappa_{n} c_n^2/2$. The solution of (\ref{rr2}) for $a_1 < 0$ is  
\begin{equation}
h=c_1\sin({\sqrt{-a_1}r})+c_2\cos({\sqrt{-a_1}r}),
\end{equation}
where the integration constants $c_1, c_2$ are required, from (\ref{ij2}),  to verify
\begin{equation}
a_1 (c_1^2 + c_2^2) +  \gamma_n=0.
\end{equation}
Since $a_1<0$, the last equation implies that this case is only compatible with a transverse section chosen as a $(n-2)$-dimensional Einstein space having a positive $k$.

In summary, we have determined all the possible solutions associated to the line element (\ref{gauge-higher-other}), where $F$ is given in (\ref{FApend}), and $G=h^{-2}$. The electric field is $F_{rt}=q /F^2$ and the scalar 
field $\phi=c \, r+\phi_0$, with $\phi_0$ an integration constant.

\end{document}